\documentclass[twocolumn,showpacs,preprintnumbers,superscriptaddress,amsmath,amssymb,prl]{revtex4}

\usepackage{amssymb}
\usepackage{stmaryrd}
\usepackage{amsmath}
\usepackage{txfonts}
\usepackage{graphicx}
\usepackage{CJK}

\usepackage{bm}
\usepackage{ulem}

\begin{document}
\begin{CJK*}{GB}{SongMT}
\CJKfamily{gbsn}

\title{Arbitrary orbital angular momentum of photons}

\author{Yue Pan}
\affiliation{School of Physics and Key Laboratory of Weak-Light Nonlinear Photonics, Nankai University, Tianjin 300071, China}
\author{Xu-Zhen Gao}
\affiliation{School of Physics and Key Laboratory of Weak-Light Nonlinear Photonics, Nankai University, Tianjin 300071, China}
\author{Zhi-Cheng Ren}
\affiliation{School of Physics and Key Laboratory of Weak-Light Nonlinear Photonics, Nankai University, Tianjin 300071, China}
\author{Xi-Lin Wang}
\affiliation{School of Physics and Key Laboratory of Weak-Light Nonlinear Photonics, Nankai University, Tianjin 300071, China}
\author{Chenghou Tu}
\affiliation{School of Physics and Key Laboratory of Weak-Light Nonlinear Photonics, Nankai University, Tianjin 300071, China}
\author{Yongnan Li}
\affiliation{School of Physics and Key Laboratory of Weak-Light Nonlinear Photonics, Nankai University, Tianjin 300071, China}
\author{Hui-Tian Wang}
\email{htwang@nju.edu.cn/htwang@nankai.edu.cn}
\affiliation{School of Physics and Key Laboratory of Weak-Light Nonlinear Photonics, Nankai University, Tianjin 300071, China}
\affiliation{National Laboratory of Solid State Microstructures, Nanjing University, Nanjing 210093, China}
\affiliation{Collaborative Innovation Center of Advanced Microstructures, Nanjing University, Nanjing 210093, China}

\date{\today}

\begin{abstract}
\noindent
Orbital angular momentum (OAM) of photons, as a new fundamental degree of freedom, has excited a great diversity of interest, because of a variety of emerging applications. Arbitrarily tunable OAM has gained much attention, but its creation remains still a tremendous challenge. We demonstrate the realization of well-controlled arbitrary OAM in both theory and experiment. We present the concept of general OAM, which extends the OAM carried by the scalar vortex field to the OAM carried by the azimuthally varying polarized vector field. The arbitrary OAM has the same characteristics as the well-defined integer OAM: intrinsic OAM, uniform local OAM and intensity ring, and propagation stability. The arbitrary OAM has unique natures: it is allowed to be flexibly tailored and the radius of the focusing ring can have various choices for a desired OAM, which are of great significance to the benefit of surprising applications of the arbitrary OAM.
\end{abstract}



\maketitle
\end{CJK*}
\newpage

\noindent Besides spin angular momentum (SAM), a scalar (homogeneously polarized) vortex field with a helical phase of $\exp (jm\phi)$ can also carry an intrinsic and eigen orbital angular momentum (OAM) of $m \hbar$ per photon \cite{R01}. The photon OAM, as a fundamental controlling degree of freedom and infinite quantum states of photons, has intrigued a broad interest \cite{R02,R03}, due to its important applications in many realms \cite{R04,R05,R06,R07,R08,R09,R10,R11,R12,R13,R14}. The photon OAM has been extended to other waves \cite{R15,R16,R17,R18}. The OAM is undoubtedly an extensively interesting topic and brings surprisingly the recent exploitations.

The fractional OAM has been given increasing attention \cite{R19,R20,R21,R22}. Differently from the scalar vortex field carrying the integer OAM, the light field carrying the fractional OAM will in general undergo an evolution during its propagation, and its local OAM and intensity pattern exhibit both the azimuthal nonuniformity and even its propagation is unstable. However, it is always anticipated that the arbitrary OAM not only is continuously tunable and propagation stable, but also has the uniformity in both the local OAM and the intensity ring, as the integer OAM. Therefore, creating and tailoring the fractional OAM meets still a tremendous challenge.
\begin{figure}[bh]
  \centering{\includegraphics[width=8.5cm]{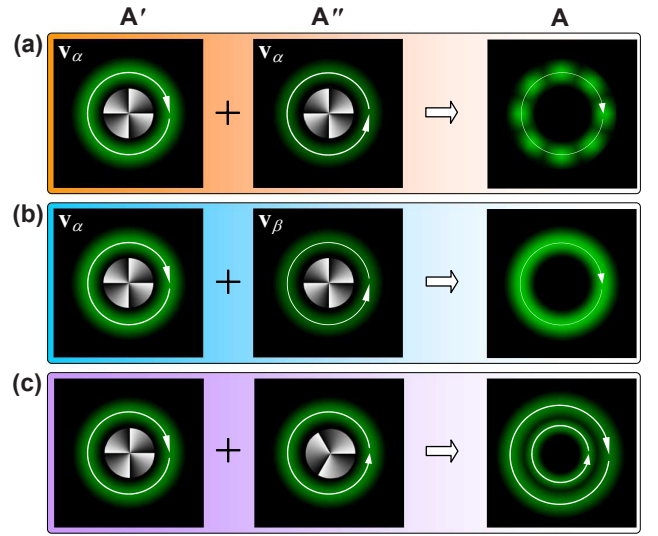}\\
  \caption{Damping model for understanding the arbitrary OAM. (a) $\mathbf{A}'$ and $\mathbf{A}''$ have the same polarization and the completely opposite OAMs of $\pm m \hbar$. (b) $\mathbf{A}'$ and $\mathbf{A}''$ have the orthogonal polarizations and the completely opposite OAMs of $\pm m \hbar$. (c) $\mathbf{A}'$ and $\mathbf{A}''$ have the OAMs of $+ m \hbar$ and $- m' \hbar$. Green patterns are schematic intensity distributions, Gray patterns are schematic vortex phases, and the circles with arrow show the sense of OAM and the lengths of arrow show the magnitudes of OAM.}}
\end{figure}

The attempts at creating the fractional OAM almost focuses only on the scalar fields. To get the substantial progress on the arbitrary OAM, the most possible solution may be to break the limit of scalar fields, by using the vector fields \cite{R23,R24,R25}. Here we report the realization of well-controlled arbitrary OAM, based on the vector fields. We present the general OAM, which extends the OAM carried by the scalar vortex field to the OAM carried by the azimuthally varying vector field (AV-VF). By using the optical tweezers, we demonstrate the arbitrary OAM, which has the same characteristics as the integer OAM: intrinsic OAM, uniform local OAM and intensity ring, and propagation stability.

\textit{Damping model}.---For the orbital motion of a classical particle, if a damping is introduced, its motion speed becomes slow and then its OAM also becomes small synchronously. This classical damping model enlightens us ``Whether introducing the suitable damping is able to flexibly realize the control of photon OAM?" A scalar vortex field $\textbf{A}'$ with the helical phase of $\exp(+jm\phi)$ is able to drive the orbital motion of the trapped particle due to the presence of OAM flux. Of course, another scalar vortex field $\textbf{A}''$ with the opposite helical phase of $\exp(-jm\phi)$ will provide the opposite-sense OAM flux, as a damping for $\textbf{A}'$. For the first consideration, $\textbf{A}''$ has the same polarization as $\textbf{A}'$ defined by the unit vector $\textbf{v}_\alpha$, thus the total field $\textbf{A} = \textbf{A}' + \textbf{A}''$ is still a scalar field. The OAM flux provided by $\textbf{A}'$ can be completely or partially canceled by the damping field $\textbf{A}''$, which is able to realize the continuously tunable net OAM flux and then the arbitrary OAM [Fig.~1(a)]. However, this is not what we ideally expected, because the interference between $\textbf{A}'$ and $\textbf{A}''$ gives rise to the nonuniformity in both the ring intensity and the local OAM [Fig.~1(a)]. Fortunately, the vector nature of photons provides a possible solution. When the polarization state of $\textbf{A}''$ defined by the unit vector $\textbf{v}_\beta$ is orthogonal to $\textbf{v}_\alpha$ of $\textbf{A}'$, the total field $\textbf{A} = \textbf{A}' + \textbf{A}''$ becomes into a vector field \cite{R23,R24,R25}. Its net OAM should also be continuously tunable [Fig.~1(b)]. In particular, it is paramount that the intensity ring and the local OAM are both uniform in the azimuthal dimension [Fig.~1(b)]. However, there is still a question why the topological charges of $\textbf{A}'$ and $\textbf{A}''$ are chosen to be completely opposite in the above. Based on the damping model, in principle $\textbf{A}''$ can has a helical phase of $\exp (-jm'\phi)$ with $m' \neq m$. In fact, such a choice is unsuitable, because the total field $\textbf{A}$ is propagation unstable ($\mathbf{A}'$ and $\mathbf{A}''$ will separate in the radial dimension, which is independent of polarization states of $\mathbf{A}'$ and $\mathbf{A}''$) [Fig.~1(c)].

\textit{Theoretical basis}.---Although the damping mode has enlightened us a possibility for realizing of arbitrary OAM by the vector fields, we have to unveil this issue in theory. We have once predicted that a vector field with the vector potential of $\mathbf{A} = A (\alpha \mathbf{v}_\alpha + \beta \mathbf{v}_\beta  ) \exp (jkz - j\omega t)$ can carry two parts of OAM flux associated with the azimuthal gradient \cite{R26}
\noindent
\begin{subequations}
\begin{align}\label{01}
J ' _z & \propto u^2 \partial \psi/ \partial \phi, \\
J ''_z & \propto u^2 \textrm{Im}( \alpha^\ast \partial \alpha / \partial \phi + \beta^\ast \partial \beta / \partial \phi),
\end{align}
\end{subequations}
\noindent where $A$ is the complex amplitude of $\textbf{A}$, as $A = u \exp (j \psi)$ by its module $u$ and its phase $\psi$. $\alpha \textbf{v}_\alpha + \beta \textbf{v}_\beta$ is a unit vector describing the distribution of polarization states of $\textbf{A}$ with $|\alpha|^2 + |\beta|^2 \equiv 1$. The unit vectors $\textbf{v}_\alpha$ and $\textbf{v}_\beta$ indicate a pair of orthogonal polarization states, represented by a pair of antipodal points on the Poincar\'{e} sphere \cite{R25,R27}. The OAM per photon can be identified as
\noindent
\begin{subequations}
\begin{align}\label{02}
L'_z & = \hbar \partial \psi/ \partial \phi, \\
L''_z & = \hbar \textrm{Im}( \alpha^\ast \partial \alpha / \partial \phi + \beta^\ast \partial \beta / \partial \phi).
\end{align}
\end{subequations}
In fact, $L'_z$ is the well-defined OAM carried by the scalar vortex field with the helical phase of $\exp (j m \phi)$, with an intrinsic and eigen OAM of $m \hbar$ per photon \cite{R01}. We call $L'_z$ as the photon OAM of the first kind. $L''_z$ is associated with the vector field. Since $L''_z$ is always zero for a scalar field because $\alpha$ and $\beta$ are independent of $\phi$, the vector field should be a unique opportunity for tailoring of  $L''_z$. With Eq.~(2), the nonzero $L''_z$ requires the polarization states to be azimuthally varying. Although the local linearly polarized vector fields \cite{R24} and the hybridly polarized vector fields \cite{R25} exhibit the azimuthally varying polarization states, $L''_z$ is still null (S1 in Ref.~28). From the above damping mode, we are aware of the importance of $L''_z$, because it might be closely related to the arbitrary OAM.
\begin{figure}[bp]
  \centering{\includegraphics[width=8.5cm]{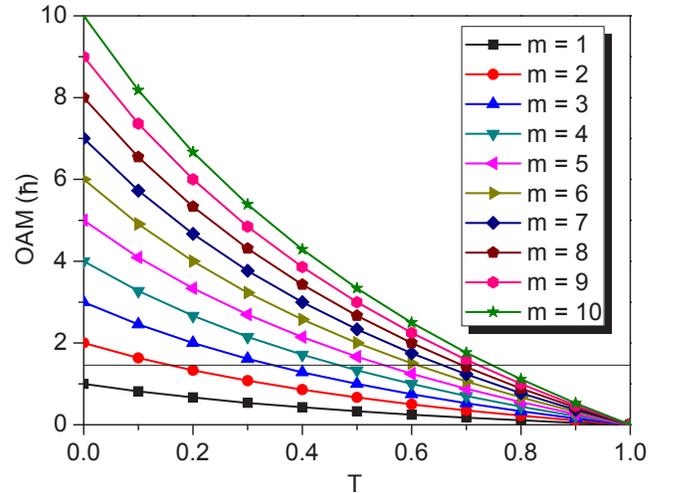}\\
  \caption{Arbitrary OAM carried by the AV-VFs. Dependence of the OAM on $m$ and $T$. Black thin horizontal line indicates a certain given OAM (or $m_{eff}$) and has a series of intersections with the color curves, in which each intersection represents a combination of $m$ and $T$ for achieving that given OAM.}}
\end{figure}

Let us refocus on the two kinds of AV-VFs \cite{R24,R25}, where a pair of orthogonally polarized components with the completely opposite helical phases of $\exp (\pm j m \phi)$ have the equal intensity. This brings us an inspiration that the most possible solution for the nonzero $L''_z$ is to break the intensity balance between the orthogonal components. The unit vector describing the distribution of polarization states can be rewritten as
\noindent
\begin{equation}\label{03}
\alpha \textbf{v}_\alpha   + \beta \textbf{v}_\beta   = \frac{1}{{\sqrt {1 + T} }}\left[ {\exp (jm\phi ) \textbf{v}_\alpha + \sqrt T \exp ( - jm\phi ) \textbf{v}_\beta  } \right],
\end{equation}
\noindent where $T$ is the relative intensity fraction between the two orthogonal components within a range of $T \in [0, 1]$. With Eqs.~(2) and (3), we have $L''_z = m \hbar (1-T)/(1+T)$, where we define an effective topological charge $m_{eff}$ as $m_{eff} = m (1-T)/(1+T)$ and the OAM per photon is $m_{eff} \hbar$. Clearly, $T$ as a degree of freedom can be indeed used to continuously tailor the OAM within a range of $[0, m \hbar]$, although $m$ can only take an integer (Fig.~2). It is very interesting and surprising that for a desired OAM or $m_{eff}$, it can be achieved by a variety of combinations of $m$ and $T$, as a series of intersections of the color curves with the thin horizontal line (Fig.~2). In the extreme case of $T = 1$, it has been confirmed $L''_z \equiv 0$ (S1 in Ref.~28). In the extreme case of $T = 0$, the vector field described in Eq.~(3) degenerates into a scalar vortex field carrying the OAM of $m \hbar$ per photon \cite{R01}. Obviously, $L'_z$ should belong to a special case of $L''_z$ when $T = 0$. In particular, the phase $\exp (j \psi)$ can be in fact incorporated into $\alpha$ and $\beta$ ($|\alpha|^2 + |\beta|^2 \equiv 1$ is still held). Therefore, $L''_z$ should be a general form of the OAM associated with the azimuthal gradient, and is called as the general OAM of the first kind and is written as $L^{(1)}_z = \hbar \textrm{Im}( \alpha^\ast \partial \alpha / \partial \phi + \beta^\ast \partial \beta / \partial \phi)$.
\begin{figure*}[th]
  \centering{\includegraphics[width=11.0cm]{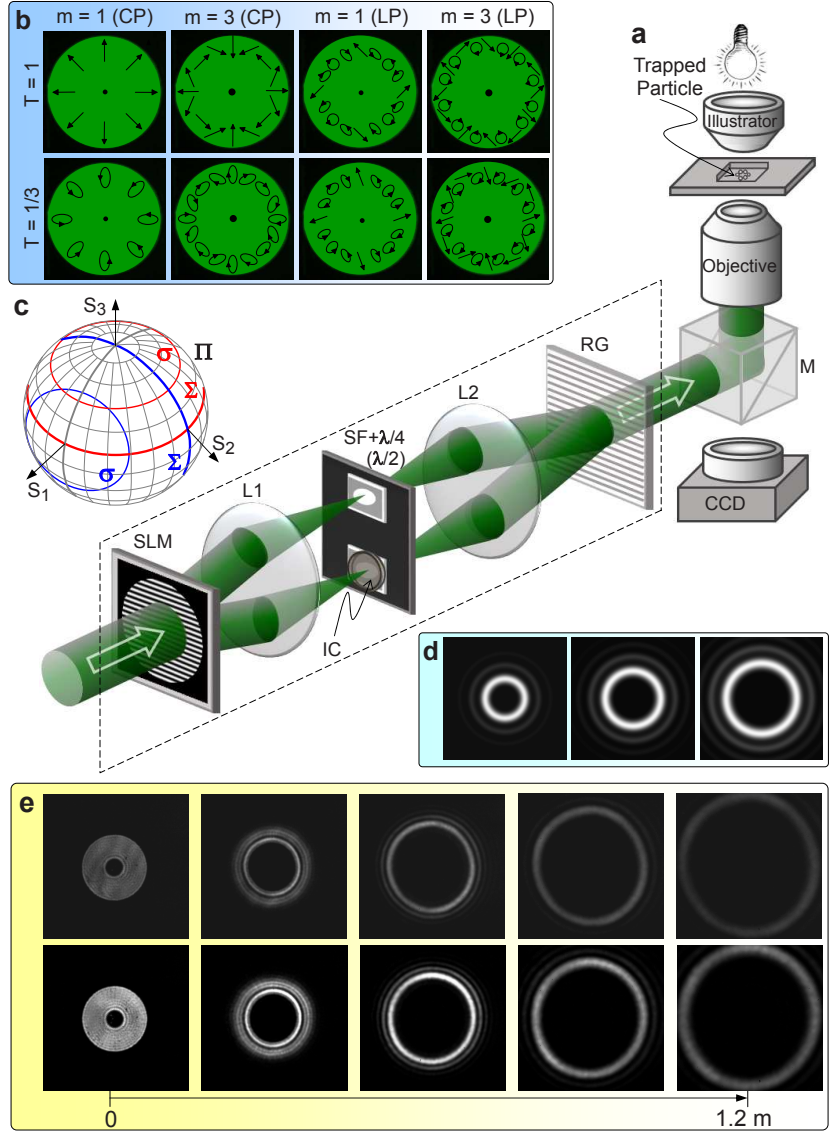}\\
  \caption{Experimental configuration to validate the arbitrary OAM. (a) Optical tweezers system, the dashed-line parallelogram shows the generation unit of AV-VFs. (b) Schematic distributions of polarization states for the AV-VFs, the 1st and 2nd (3rd and 4th) columns show the vector fields created by a pair of right- and left-handed CP ($x$ and $y$ LP) bases, the 1st and 2nd rows are the cases of $T = 1$ and $T = 1/3$, respectively. (c) Geometric presentation of the Poincar\'{e} sphere for the polarization states of the AV-VFs shown in (b). (d) Simulated multi-ring structures of the focused top-hat AV-VFs with $m =10$, 14 and 18 when $T = 1/3$. (e) Propagation evolution of the top-hat AV-VFs created by a pair of orthogonal polarized bases with $\exp (\pm j 20 \phi)$ when $T = 0.32$ in free space (top row), for comparison, the measured pattern evolution of the top-hat scalar vortex field with $\exp (+ j 20 \phi)$ is also shown (top row).}}
\end{figure*}

\textit{Experimental evidence}.---To confirm the feasibility of the arbitrary OAM, the optical tweezers is a useful tool. In the experimental schematic of the optical tweezers [Fig.~3(a)], the generation unit of the vector fields is very similar to that used in Refs.~24 and 25, but has a unique difference that the $\pm 1$st orders with the completely opposite helical phases of $\exp (\pm j m \phi)$ can have the different intensity (Methods in Ref.~28). Thus the demanded vector field can be written as
\begin{equation}\label{04}
\textbf{A} =u(r) \frac{1}{{\sqrt {1 + T} }}\left[ {\exp (jm\phi ) \textbf{v}_\alpha   + \sqrt T \exp ( - jm\phi ) \textbf{v}_\beta  } \right].
\end{equation}
\noindent where $u(r)$ has the top-hat profile with $u(r) = U_0 \textrm{circ} (r/R_0)$ [Fig.~3(b)]. $U_0$ is constant amplitude, and $\textrm{circ} (\cdot)$ is a well-known circular function (Methods in Ref.~28).

The vector fields created by a pair of orthogonal circularly polarized (CP) spinors [$\{\textbf{v}_\alpha, \textbf{v}_\beta  \}  \to \{ \textbf{v}_r, \textbf{v}_l \}$ in Eq.~(4)], as shown in the 1st and 2nd columns of Fig.~3(b), the azimuthally varying linear polarizations for $m = 1$ $(m = 3)$ when $T = 1$ traverses once (thrice) all points located at the equator on the Poincar\'{e} sphere $\Pi$ (Ref.~27, S2 and Fig.~S1 in Ref.~28) with the Stocks parameter of $s_3 = 0$ shown by the thick red curve in Fig.~3(c). The elliptical polarizations with the same ellipticity but azimuthally varying orientation for $m = 1$ $(m = 3)$ when $T = 1/3$ traverse once (thrice) all points located at the north-latitude $30^\circ$ circle $(s_3 = 1/2)$ on $\Pi$, shown by the thin red curve in Fig.~3(c). In contrast, for the vector fields created by a pair of orthogonal linearly polarized (LP) bases [$\{ \textbf{v}_\alpha, \textbf{v}_\beta  \}  \to \{ \textbf{v}_x, \textbf{v}_y \}$ in Eq.~(4)] in the 3rd (4th) column of Fig.~3(b), the polarization state undergoes the azimuthal variation from the linear, through elliptic to circular polarizations when $T = 1$, which traverse once (thrice) all points located at the great circle $(s_1 = 0)$ on $\Pi$ for $m =1$ $(m = 3)$, shown by the thick blue curve in Fig.~3(c); while the polarization state undergoes the azimuthal change from the linear to elliptic polarizations but does not occur the circular polarization when $T = 1/3$, which traverse once (thrice) all points located at the $s_1 = 1/2$ circle on $\Pi$ for $m =1$ $(m = 3)$, shown by the thin blue curve in Fig.~3(c). For a more general case (S2 and Fig. S2 in Ref.~28), a pair of orthogonally polarized bases $\{ \textbf{v}_\alpha, \textbf{v}_\beta  \}$ in Eq.~(4) correspond to any pair of antipodal points on $\Pi$. Thus the polarization states of the created vector field are described by all points located at a circle $\sigma$ on $\Pi$. The circle $\sigma$ is the intersection of $\Pi$ with the plane $\sigma$ normal to the connecting line between the antipodal points. The plane $\sigma$ has a distance of $d = (1-T)/(1+T)$ from the center of $\Pi$. We further define a great circle $\Sigma$, which is the intersection of $\Pi$ with a plane passing the center of $\Pi$ and being parallel to the plane $\sigma$. As a result, the Poincar\'{e} sphere can also characterize the arbitrary OAM, which is equal to the distance $d$ of the plane $\sigma$ from the center of $\Pi$, in units of $m \hbar$. Of course, the OAM can also be characterized as $m \hbar (\Omega / 2 \pi)$, by a solid angle $\Omega$ subtended by the spherical zone sandwiched between the two circles $\sigma$ and $\Sigma$ on $\Pi$, with $\Omega = 2 \pi (1-T)/(1+T)$ (S2 in Ref.~28).

The simulated patterns of the focused vector fields (for three different $m = 10$, 14, 18 with the same $T = 1/3$) by an objective with $NA = 0.75$ exhibit the uniform multi-ring structure composed of a principal ring and secondary rings [Fig.~3(d)]. For a given objective, the discrete radius of the principal ring depends only on $m$ independent of $T$. It is of great importance to explore the propagation stability of the vector fields carrying the arbitrary OAM. The measured intensity pattern of the top-hat scalar vortex field with $\exp (+j 20 \phi)$ undergoes an evolution from the top-hat profile at $z = 0$ to the multi-ring structure at $z = 1.2$ m [top row in Fig.~3(e)]. For the vector field created by a pair of orthogonal polarized bases with the opposite helical phases of $\exp (\pm j 20 \phi)$ when $T = 0.32$, its propagation behavior has no difference from the top-hat scalar vortex field, implying that the vector field is propagation stable [bottom row in Fig.~3(e)]. For the vector field created by a pair of orthogonal polarized bases with $\exp (+j 20 \phi)$ and $\exp (- j 5 \phi)$ when $T = 0.32$, this vector field is unstable during its propagation (S3 and Fig.~S3 in Ref. 28).
\begin{figure}[thb]
  \centering{\includegraphics[width=8.5cm]{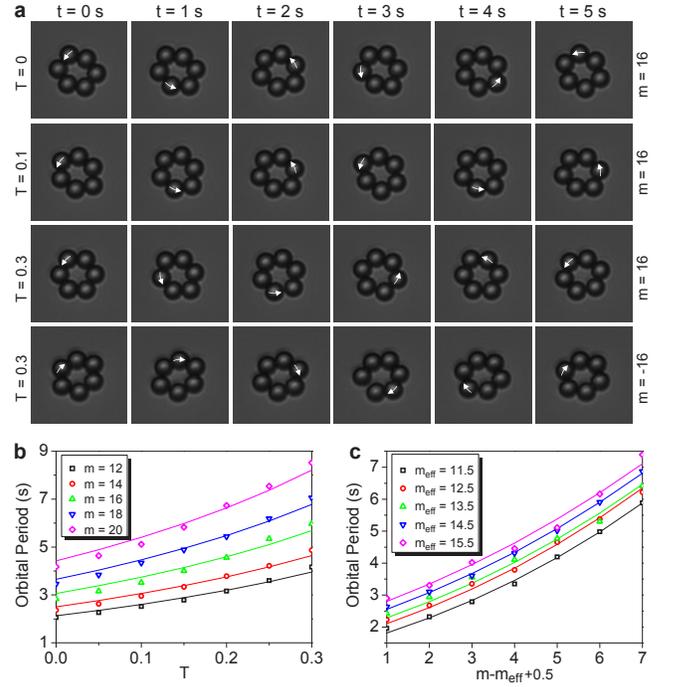}\\
  \caption{(color online)  Observed orbital motion of trapped particles around the ring focus produced by the AV-VFs. (a) Snapshots of orbital motion of trapped particles (Movie in Ref.~28). (b) Dependence of the orbital period $\tau$ on $T$ for five different $m$, and symbols are the measured periods and the corresponding curves are the fitting results obtained by $\tau \propto 1/m_{eff} = (1 + T)/(1 - T)$ (S4 in Ref.~28). (c) Dependence of the orbital period $\tau$ on $m$ for five different $m_{eff}$, and symbols are the measured periods and the corresponding curves are the fitting results obtained by $\tau \propto R^3  \propto m^3$ (S4 in Ref.~28).}}
\end{figure}

The created AV-VF is incident into the optical tweezers system (Methods in Ref.~28) to explore the photon OAM by observing the orbital motion of the trapped particles (Movie in Ref.~28). We intercept the time-lapse photos of the orbital motion of the trapped particles [Fig.~4(a)]. For the vector field with $m = 16$ and $T = 0$ (it degenerates into the scalar vortex field with $m = 16$), the trapped particles move around the principal ring focus with an orbital period $\tau$$\sim$$2.47$ s (1st row). When $T$ is changed to $T = 0.1$, $\tau$ increases to $\tau$$\sim$$2.94$ s (2nd row). When $T$ is further increased to $T = 0.3$, $\tau$ further increases to $\tau$$\sim$$4.75$ s (3rd row). When $m$ is switched from $m = 16$ to $m = -16$ when keeping $T = 0.3$, the motion direction of the trapped particles is synchronously reversed with $\tau$$\sim$$4.99$ s (4th row and Movie). However, for the vector fields with $T = 1$ (the hybridly polarized vector fields reported in Ref.~25), no orbital motion of the trapped particles is observed, implying that such a kind of vector fields carry no OAM. The dependences of $\tau$ on $T$ for different $m$ [Fig.~4(b)] and on $m$ for different $m_{eff}$ [Fig.~4(c)] indicate that the measured orbital periods (symbols) are in good agreement with the fitted curves by the respective formulae $\tau  \propto m^2 (1 + T)/(1 - T)$ and $\tau  \propto m^3 /m_{eff}$ (S4 in Ref.~28). The observed results clarify the following facts. (i) The AV-VFs could indeed carry the arbitrary OAM and (ii) the choice of orthogonally polarized bases has no influence on the arbitrary OAM.

\textit{Summary}.---We present a realization of the arbitrary OAM for long-time challenge. It has novel and unique natures: (i) it is continuously tunable within a range of $[- m \hbar, m \hbar]$, (ii) it has the uniformity, and (iii) the light field carrying the arbitrary OAM has the uniform intensity ring and has the propagation stability. We also present the general OAM of the first kind, associated with the azimuthal gradient, which extends the OAM carried by the scalar vortex fields to the OAM carried by the AV-VFs. We also extend the Poincar\'{e} sphere to represent the arbitrary OAM. Our idea may spur further independent insights into the generation of natural waves carrying the arbitrary OAM. The current technology trend has been perceived to direct from fundamental investigations towards probing its viabilities for surprising applications. The fast-moving exploitation on such diverse areas has pushed for further development on OAM generation technology.

This work is supported by the 973 Program of China under Grant No. 2012CB921900, the National Natural Science Foundation of China under Grant Nos. 11534006, 11274183 and 11374166, and the National scientific instrument and equipment development project 2012YQ17004.


\end{document}